\documentclass[journal]{IEEEtran}
\usepackage[pdftex]{graphicx}
\usepackage{array}
\usepackage{mdwmath}
\usepackage{mdwtab}
\usepackage{fixltx2e}
\usepackage{stfloats}
\usepackage{color}
\usepackage{ulem}

\usepackage{url}

\begin{document}
\title{Tackling the blackbody shift \\
in a strontium optical lattice clock}
\author{
Th. Middelmann\thanks{Email: thomas.middelmann@ptb.de}, Ch. Lisdat, St. Falke, J.S.R. Vellore Winfred, F. Riehle, and U. Sterr\\
Physikalisch-Technische Bundesanstalt, Bundesallee 100,
38116 Braunschweig, Germany}
\maketitle
\begin{abstract}
A major obstacle for optical clocks is the frequency shift due to black body radiation. We discuss how one can tackle this problem in an optical lattice clock; in our case $^{87}$Sr: firstly, by a measurement of the dc Stark shift of the clock transition and, secondly, by interrogating the atoms in a cryogenic environment. Both approaches rely on transporting ultracold atoms over several cm within a probe cycle. We evaluate this approach of mechanically moving the optical lattice and conclude that it is feasible to transport the atoms over 50~mm within 300~ms. With this transport a dc Stark shift measurement will allow to reduce the contribution of the blackbody radiation to the fractional uncertainty below $2~\times~10^{-17}$ at room temperature by improving the shift coefficient known only from atomic structure calculations up to now. We propose a cryogenic environment at 77~K that will reduce this contribution to few parts in $10^{-18}$.

\end{abstract}


%
\IEEEpeerreviewmaketitle

\section{Introduction}
%
%
%
Optical clocks can achieve a higher stability and lower systematic uncertainty than current microwave clocks using a hyperfine transition of $\mathrm{^{133}}$Cs that defines the SI second. 
In an optical clock an optical atomic transition serves as absolute reference which is interrogated either in an ensemble of neutral atoms or in a single ion. 
The effect of blackbody radiation from the environment of the atoms is one of the main limitations of the systematic uncertainty of optical clocks, in particular for neutral Sr optical clocks. The radiation in thermodynamic equilibrium with the environment shifts each clock level in energy by coupling it to other states \cite{ita82}. In $^{87}$Sr the difference of the two shifts results in a line shift of about 2.4~Hz at room temperature \cite{por06}. 
In the best frequency measurement of the Sr clock transition insufficient knowledge of the shift is contributing with about $\mathrm{1\times10^{-16}}$, which is the largest contribution to the systematic uncertainty \cite{cam08b}.
Up to now the value of the blackbody shift is derived from atomic structure calculations \cite{por06} and no measurements have been published for strontium so far. 

Here we present two experimental approaches to measure or reduce the Sr blackbody shift in order to reduce its fractional uncertainty contribution to the clock below $\mathrm{10^{-17}}$. 

We first provide a description of our experimental setup and the sequences used to trap and interrogate $\mathrm{^{87}}$Sr (section~\ref{sec:expseq}).
Then we give a short introduction to the blackbody shift (section~\ref{sec:gen}) and present the two experimental approaches. First, we describe how to conduct a dc Stark shift measurement of the Sr clock transition (section~\ref{sec:dcstark}).
Second, we discuss how to measure the Sr clock transition frequency in a cryogenic environment (section~\ref{sec:lowt}). 
An interrogation of the atoms in a well controlled electric field or a low temperature environment respectively is possible only outside the magneto-optical trap (MOT) where good optical access is needed. We will use a moving lattice to transport the atoms over 50~mm. Tests of the performance of our moving lattice setup are presented in section~\ref{sec:movlat}.

Both approaches have been successfully employed for $\mathrm{^{133}}$Cs primary standards \cite{sim98,lev10}. 
It is possible to reduce the fractional uncertainty due to the blackbody shift in room temperature Cs fountain clocks to below $\mathrm{10^{-16}}$, such that it is not the limiting contribution any more. The blackbody shift itself is known with a fractional uncertainty of  $3.6 \times 10^{-3}$ \cite{ger10}.

\section{Experimental Setup}
\label{sec:expseq}
Our experimental setup has the capability to prepare either $^{88}$Sr or $^{87}$Sr in an optical lattice to perform high resolution spectroscopy on the clock transition $^3$P$_0 - ^1$S$_0$ at 698~nm. Since we have described the cooling and spectroscopy sequence for $^{88}$Sr before \cite{leg09, lis09} we will focus on changes of the experimental details with regard to the spectroscopy of $^{87}$Sr \cite{lis10}.

A cold atomic beam is generated by Zeeman-slowing a thermal atomic beam evaporated from a furnace of about 480\,$^{\circ}$C. The slowed atomic beam is deflected by a 2D optical molasses into a part of the vacuum system with no direct line of sight on the oven, where a MOT is operated. The coils generating the magnetic quadrupole field and later a homogeneous field are located inside the vacuum chamber. 

First the atoms are trapped and cooled on the 461~nm transition $^1$P$_1 - ^1$S$_0$. The cooling laser is stabilized to an ultra-stable resonator to achieve short- and long-term frequency stability. During the first MOT phase we apply repumping laser light on the transitions $^3$S$_1 - ^3$P$_0$ (679~nm) and $^3$S$_1 - ^3$P$_2$ (707~nm). Due to the nuclear spin of $I=9/2$ of $^{87}$Sr and the connected hyperfine structure several frequencies are required to fully remove the population in the $^3$P$_2$ level. We achieve this by rapidly scanning the 707~nm repump-laser over an interval of about 5~GHz. Sufficient long-term stability of the frequency is achieved by stabilizing the average frequency of the laser with a commercial wavemeter, while the unmodulated repump-laser at 679~nm is stabilized to an ultra stable cavity.

Having cooled the strontium atoms to about 3~mK the first MOT stage is replaced by a second MOT stage on the 689~nm intercombination line $^3$P$_1 - ^1$S$_0$. To enlarge the capture velocity of the second MOT stage, a frequency modulation of the cooling laser is applied. The cooling on the intercombination line of $^{87}$Sr is hampered by strongly different $g$-factors in the upper and lower state \cite{muk03}. To enhance the cooling efficiency we apply -- in addition to the cooling light on the $F'=11/2 - F = 9/2$ transition -- so called stirring light on the $\Delta F = 0$ hyperfine transition. Light for this purpose is generated by phase-locking a second extended cavity diode laser at the appropriate frequency difference to the master laser for the cooling on the intercombination line. The stirring laser beams are superimposed with the cooling beams and have the same polarization.
\begin{figure}[!t]
\centering
\includegraphics[width=\columnwidth]{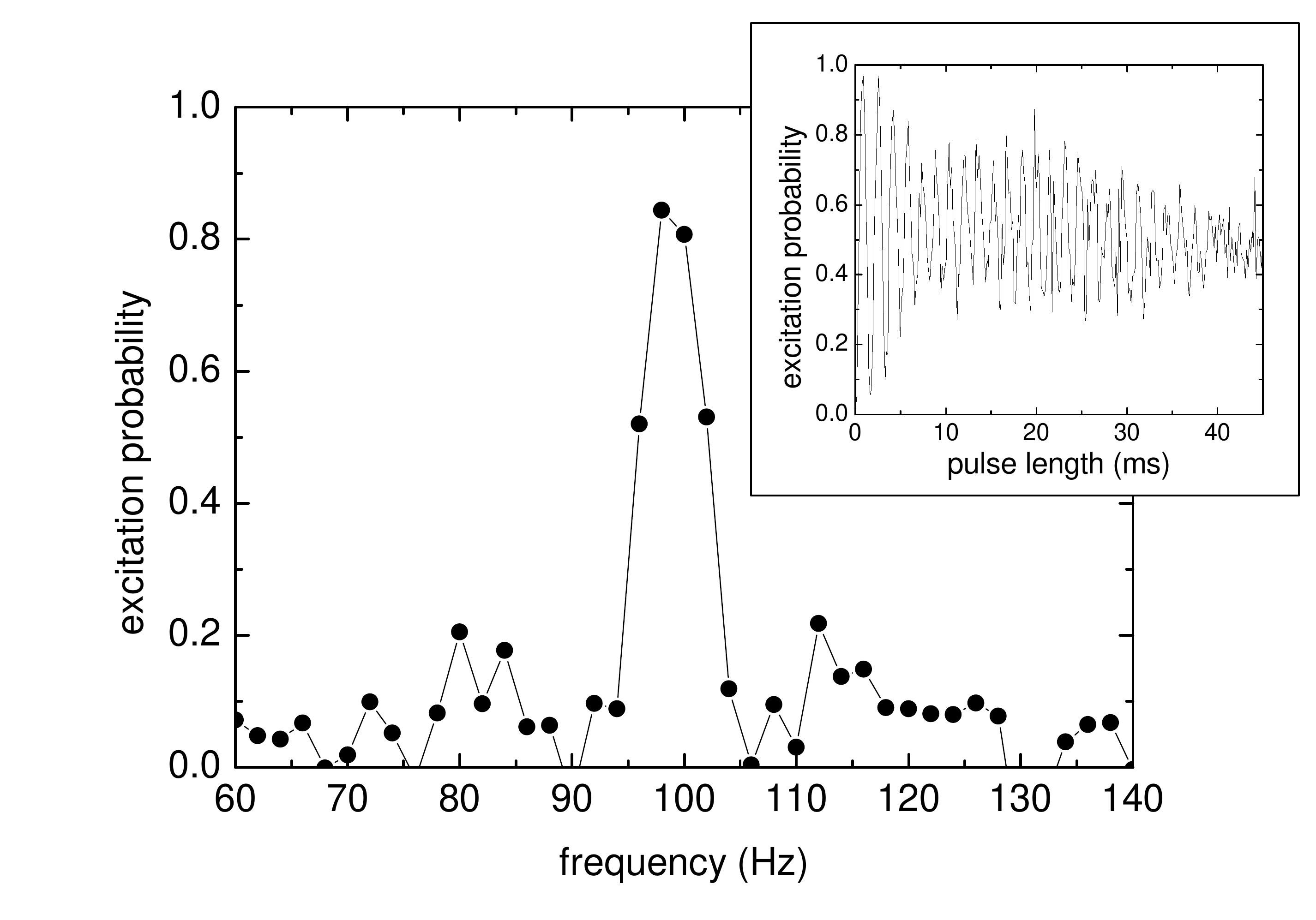}
\caption{Spectrum of the highest frequency Zeeman component of the $^{87}$Sr $^3$P$_0 - ^1$S$_0$ clock transition with a Fourier-limited linewidth of 10~Hz. The zero point of the frequency axis is roughly at the transition frequency without magnetic field. The inset shows Rabi oscillations recorded at higher pulse power. The revival of the signal is due to beating of Rabi oscillations from atoms in different vibrational levels of the optical lattice.}
\label{fig:scan}
\end{figure}
Applying the frequency modulation of the cooling and stirring light we achieve a temperature of the atoms of about 50~$\mu$K. The temperature is further reduced to about 4~$\mu$K by switching off the frequency modulation and the stirring laser beams and lowering the laser beam intensity. During this last cooling phase the atoms are efficiently loaded into a horizontally oriented 1D optical lattice at the magic wavelength \cite{kat03}. Light is provided by a Ti:Sapphire laser coupled to a single mode fiber. At maximum, 600~mW are available at the fiber output. Usually, the experiment is operated with 150~mW focused to a $1/e^2$ waist radius $w_0$~=~32~$\mu$m at the position of the atoms. This beam is retro-reflected to obtain the lattice. To load about 10$^5$ $^{87}$Sr atoms we require typically 200~ms, 90~ms, and 50~ms for the three cooling phases.

Before we perform high resolution spectroscopy we spin-polarize the atoms to either the $m_F = +9/2$ or $m_F = -9/2$ Zeeman sublevel. For this purpose a weak homogeneous magnetic field of about 22~$\mu$T is applied and the atoms are irradiated for 1~ms by resonant light from the stirring laser with either right- or left-handed circular polarization.

To remove population in the undesired Zeeman levels we increase the magnetic field to about 1.8~mT and apply a high intensity short $\pi$-pulse of typically 1~ms length resonant with the desired Zeeman transition. The population remaining in the $^1$S$_0$ level is then removed from the lattice by resonant radiation on the 461~nm line. Afterwards, the magnetic field is lowered again to 22~$\mu$T; typically $2 \times 10^4$ atoms remain. This field is large enough to remove the degeneracy of the Zeeman transitions and minimize line pulling effects. The clock transition is interrogated by a 90~ms $\pi$-pulse that produces 10~Hz wide Fourier-limited lines with high contrast (Fig.~\ref{fig:scan}). For detection, first the $^1$S$_0$ ground state population is detected by laser induced fluorescence of the $^1$P$_0 - ^1$S$_0$ transition and then removed by the radiation pressure. The population in the upper clock state is similarly detected after repumping it to the ground state. In this way the excitation probability is measured independent from atom number fluctuations. 
\begin{figure}[!t]
\centering
\includegraphics[width=\columnwidth]{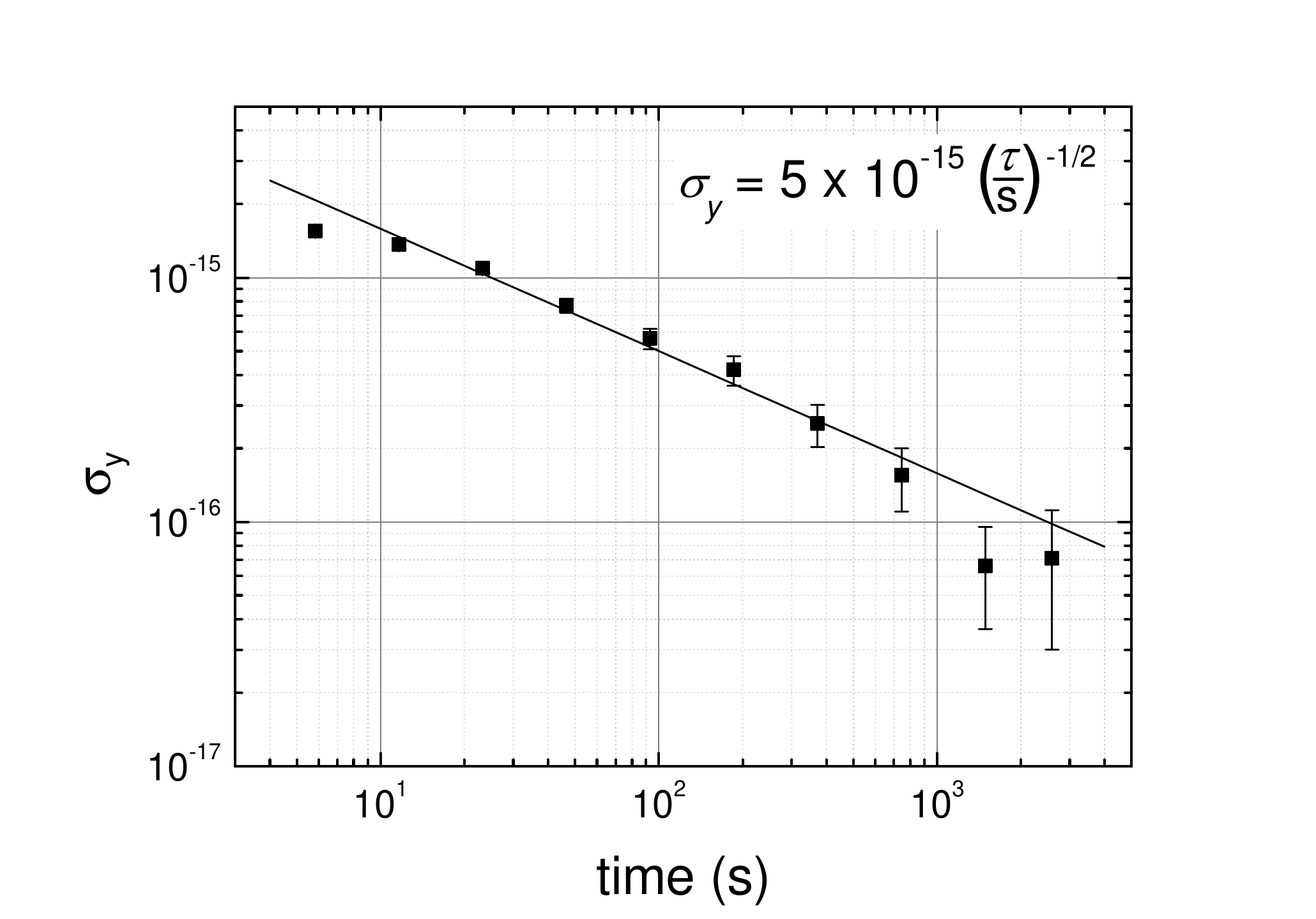}
\caption{Allan deviation of a signal from an interleaved stabilization of the Sr clock laser to the $^3$P$_0 - ^1$S$_0$ clock transition. The functional dependence for white frequency noise is indicated by the  straight line.}
\label{fig:allan}
\end{figure}

To stabilize the 698~nm spectroscopy laser to the atoms, the two outermost Zeeman components of the $^3$P$_0 - ^1$S$_0$ transition are interrogated alternatingly on their high- and low-frequency half-maximum points. Using the four observed excitation probabilities, the Zeeman splitting is determined and the laser is steered to the magnetic field-free frequency by variation of the offset frequency between laser and reference cavity. 

For the investigation of systematic effects, we typically use two interleaved stabilization cycles, in which we modulate the effect to be investigated. The mean difference of the offset frequencies of two consecutive stabilization cycles is the desired frequency shift. Other effects like the cavity drift are suppressed to a high degree. The Allan deviation of an interleaved stabilization signal is plotted in Fig.~\ref{fig:allan}. It scales as $\tau^{-1/2}$, crossing $10^{-16}$ at a few 1000~s. Thus shifts
can be determined with a relative uncertainty of $10^{-16}$ within a few 1000~s averaging time.

\section{Blackbody radiation shift}
\label{sec:bbr}
\subsection{General}
\label{sec:gen}

For Sr the bulk spectral distribution of the room temperature blackbody radiation (BBR) is at much lower frequencies than the relevant atomic transitions.
In this case the  induced shifts can be described by static shifts due to the average (rms) electric field plus small dynamic corrections \cite{far81,deg05a}. The energy shift of the level $i$ due to blackbody radiation is given by 
\begin{equation}
\label{eq:bbrmain}
\delta E_i({T})  = \delta {E^{\rm(stat)}_{i}}({T}) + \delta E^{\rm(dyn)}_{i}({T}).
\end{equation}
The static contribution
\begin{equation}
\label{eq:bbrstat}
\delta {E^{\rm(stat)}_{i}}({T}) =  -\frac{{\alpha}_{i}}{2} \langle E^2 \rangle                           
\end{equation}
depends on the static polarizability $\alpha_i$ and the mean square of the electric field $\langle E^2 \rangle$ at temperature $T$ which is calculated via the radiant energy density $\rho$ according to the Stefan-Boltzmann law
\begin{equation}
\label{eq:efield}
\langle E^2 \rangle = \frac{\rho}{\epsilon_0} = \frac{8\pi^5 k_B^4}{15 c_0^3 \epsilon_0
h^3} \;{T}^4=c_{1}\cdot {T}^4.
\end{equation}
with the constant ${c_{1}}=8.545\cdot 10^{-5}\frac{\mathrm{V}^2}{\mathrm{m}^2\mathrm{K}^4}$.
The dynamic part
\begin{eqnarray}
\label{eq:bbrdyn}
\delta {E^{\rm(dyn)}_{i}}({T}) & = & - \frac{\hbar}{2 \pi} \left(\frac{k_B {T}}{\hbar}\right)^3 \nonumber\\
&& \ \ \times \sum_k \frac{2 J_k + 1}{2 J_i +1} \frac{A_{ki}}{\omega_{ik}^3} \ \ G\left(\frac{\hbar\omega_{ik}}{k_B T}\right)
\end{eqnarray}
contains the contributions from all dipole transitions from the considered state
$i$ to levels $k$ with the respective Einstein coefficients
$A_{ki}$ (spontaneous emission rate for $E_k > E_i$), and transition frequencies
$\nu_{ik}=\omega_{ik}/2\pi$ weighted by the function 
\begin{eqnarray}
G(y) &=& \int_0^\infty\frac{x^3}{e^x-1}\left(\frac{1}{y-x}+\frac{1}{y+x}-\frac{2}{y} \right)\,dx \nonumber \\
     & \approx& \frac{16 \pi^6}{63 y^3}{\rm \ \ \ for \ } y >> 1 .
\end{eqnarray}
The frequency shift of the clock transition due to blackbody radiation 
$\Delta\nu_{\rm {BBR}}=\Delta\nu^{\rm(stat)}+\Delta\nu^{\rm(dyn)}$ 
can be expressed as a sum of the static Stark shift
\begin{equation}
\label{eq:bbrSstat}
\Delta {\nu^{\rm(stat)}}({T}) = -\frac{\Delta\alpha}{2 h} \langle E^2 \rangle =- \frac{c_{1}}{2h} \; \Delta\alpha {T}^4                     
\end{equation}
and a dynamic correction
\begin{equation}
\label{eq:bbrSdyn}
\Delta {\nu^{\rm(dyn)}}({T}) =  [{\delta {E^{\rm(dyn)}_{e}}({T}) - \delta {E^{\rm(dyn)}_{g}}({T})}]/{h}                           
\end{equation}
where $g$ and $e$ denote the ground and excited state of the clock transition, respectively.
With the difference of the polarizabilities of the two clock states 
$\Delta\alpha=\alpha_e-\alpha_g={\rm 4.304(59)\times10^{-39}}$~Cm$^2/$V~\cite{por06} 
the static contribution is
$\Delta {\nu^{\rm(stat)}}({\rm 300\:K})$~=~2.248(31)~Hz.
The dynamic contribution of the $\mathrm{^{1}S_{0}}$ state is negligible. Using the data given in ~\cite{por06} we calculate the dynamic contribution of the $\mathrm{^{3}P_{0}}$ state and obtain
$\Delta {\nu^{\rm(dyn)}}({\rm 300\:K})=0.107(11)$~Hz.
The 10\% uncertainty of the dynamic correction is estimated from the one of the Einstein coefficient of the 5s4d $^3$D$_1 - ^3$P$_0$ transition, which contributes to more than 90\% to the dynamic part. We have inferred the uncertainty of the Einstein coefficient from half the difference between the measured and calculated radiative life time of the 5s4d $^3$D state as discussed in \cite{por08}.

The sum of both values is given as $\Delta\nu_{\rm {BBR}}({\rm 300\:K})=2.354(32)$~Hz \cite{por06}. 
Comparing the uncertainties of the static and the dynamic part it is obvious that the main contribution to the Sr blackbody shift uncertainty arises currently from the knowledge of $\Delta\alpha$.

\subsection{DC Stark shift}
\label{sec:dcstark}
\begin{figure}[!t]
\centering
\includegraphics[width=0.6\columnwidth]{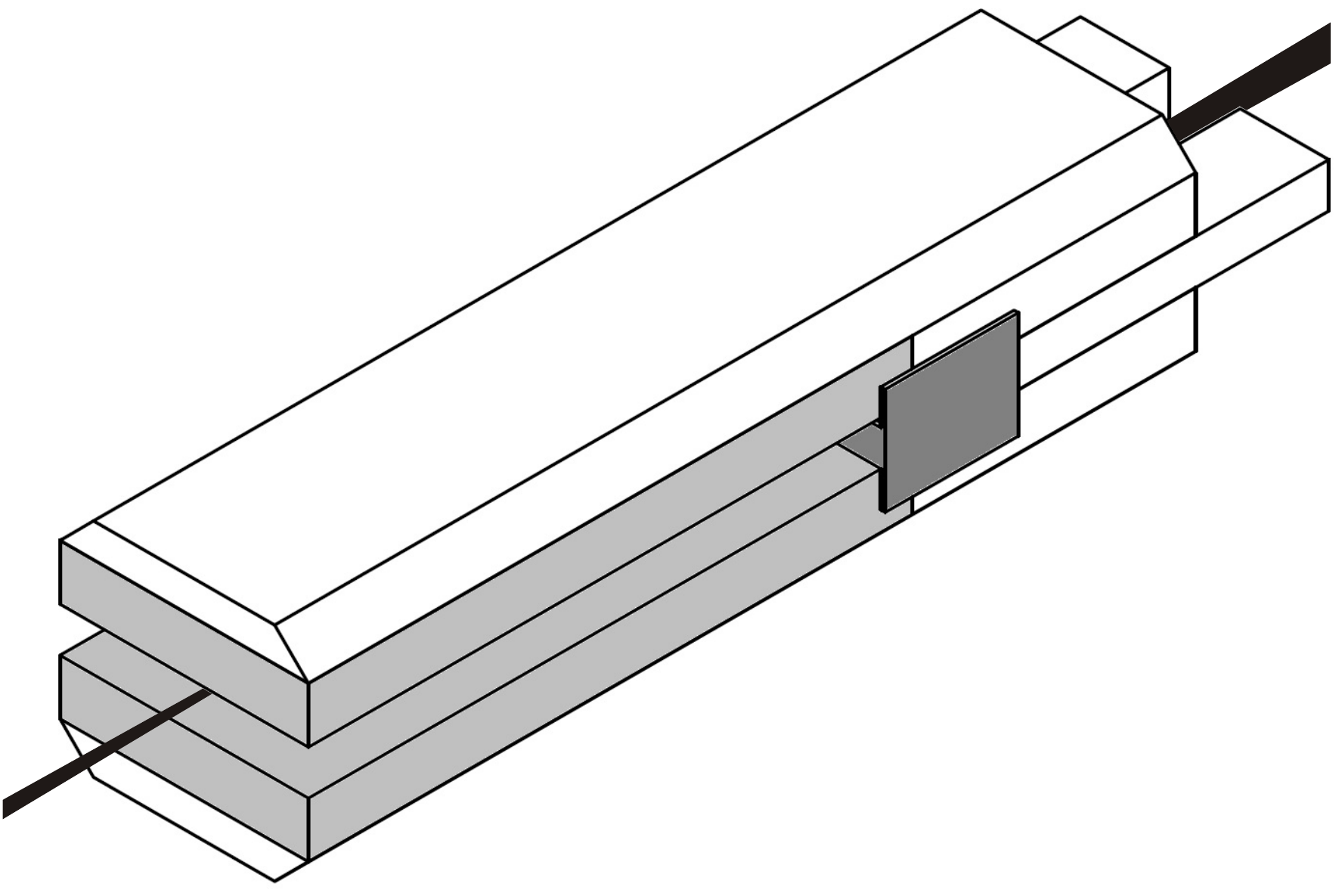}
\caption{Sketch of the capacitor. The lattice laser beam is depicted in black, the gold-coated conducting surfaces in light gray, and the screening shields in dark gray. The field plate area is 30~mm $\times$ 70~mm.}
\label{fig:capa}
\end{figure}

To tackle the main uncertainty contribution $U(\Delta\alpha)$ we plan to perform a dc Stark shift measurement 
by interrogating  atoms in a homogeneous electric field. It will be created by two parallel field plates (Fig.~\ref{fig:capa}) inducing a frequency shift large enough for a measurement with sufficient accuracy.
We plan to use field plates from Zerodur (low thermal expansion glass-ceramic) coated with a thin semi-transparent gold layer. The needed precision of separation and parallelism can be realized by using 5~mm Zerodur gauge blocks as spacer. Interferometric methods will be applied for verification \cite{win84a}. 
The individual Zerodur parts will be connected by optical contact bonding.

The design is such that all dielectric surfaces are as far as possible from the position of interrogation. Surface charges on the spacers will be screened by conducting shields (Fig.~\ref{fig:capa}), held at the average potential and separated by a small distance from the electrodes. 

\begin{figure}[!t]
\centering
\includegraphics[width=0.95\columnwidth]{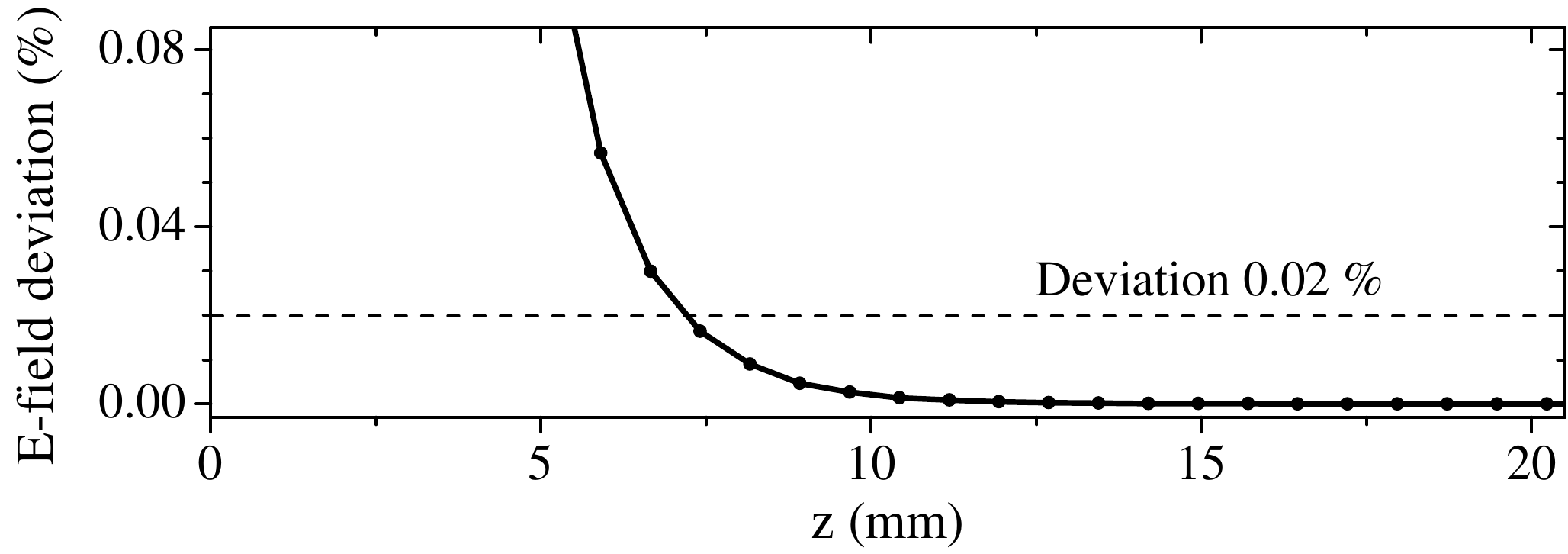}
\caption{Finite elements simulation of the electric field component perpendicular to the field plates along the symmetry axis (c.f. Fig~\ref{fig:capa}). The curve shows the deviation of the electric field from the one of an infinite capacitor as a function of the distance from the edge of the field plates.}
\label{fig:esim}
\end{figure}

We want to implement the field plate arrangement in our existing setup and thus must meet tight spatial constraints.
For our suitable design we have checked the field homogeneity with a finite elements calculation.
Figure~\ref{fig:esim} shows the homogeneity of the electric field. For a penetration depth of about 15~mm along the central axis between the two field plates it is better than $2\times 10^{-4}$, which poses no limit on our measurement.
In order to estimate the achievable uncertainty of the polarizability difference
\begin{equation}
\label{eq:alsta}
\Delta\alpha = -2h\Delta\nu\frac{d^2}{U^2}
\end{equation}
we assume fractional uncertainties better than $2\times 10^{-4}$ for the spacing of the field plates $d$, the applied voltage $U$, and the induced frequency shift $\Delta\nu$. With this we expect a fractional uncertainty of about $7\times 10^{-4}$, which is an improvement by a factor of 20 compared to the value from atomic structure calculations mentioned above \cite{por06}.

With this uncertainty $\Delta\alpha$ is contributing with only $\mathrm{4\times 10^{-18}}$ to the systematic uncertainty of the determination of the unperturbed Sr transition. The uncertainty of the blackbody shift is then limited by the knowledge of the dynamic correction to $\mathrm{2.5\times10^{-17}}$.
The uncertainty of the correction may be further reduced by refining the atomic structure calculations with the new value for $\Delta\alpha$ \cite{por08}.

\subsection{Frequency measurement at low temperature}
\label{sec:lowt}

It is important to note that so far we have not discussed the uncertainty arising from the temperature knowledge of the surroundings, stemming from temperature gradients and temperature measurements. A temperature uncertainty  of e.g. 0.1~K at 300~K would cause a fractional uncertainty of the clock frequency of $\mathrm{7\times10^{-18}}$.

In a complementary approach the BBR shift can be largely reduced by operating the clock at low temperature, as suggested in \cite{lud08}.
At liquid nitrogen temperature the BBR shift is $\Delta {\nu_{\rm {BRR}}}({\rm 77\:K})=0.01$~Hz, which corresponds to $2.4\times 10^{-17}$ in the clock frequency. With the atomic response as known from literature and a temperature uncertainty of $U(T)=1$~K, the uncertainty contribution arising from  77~K BBR is $<2\cdot 10^{-18}$.

In this experiment we consider a cold cavity with a small orifice through which atoms may enter or leave. We have to take the effect of BBR photons entering through this hole into account. 
To achieve a low uncertainty of the blackbody shift it is crucial to keep the disturbance due to this incoming room temperature BBR small.

\begin{figure}[!t]
\centering
\includegraphics[width=0.8\columnwidth]{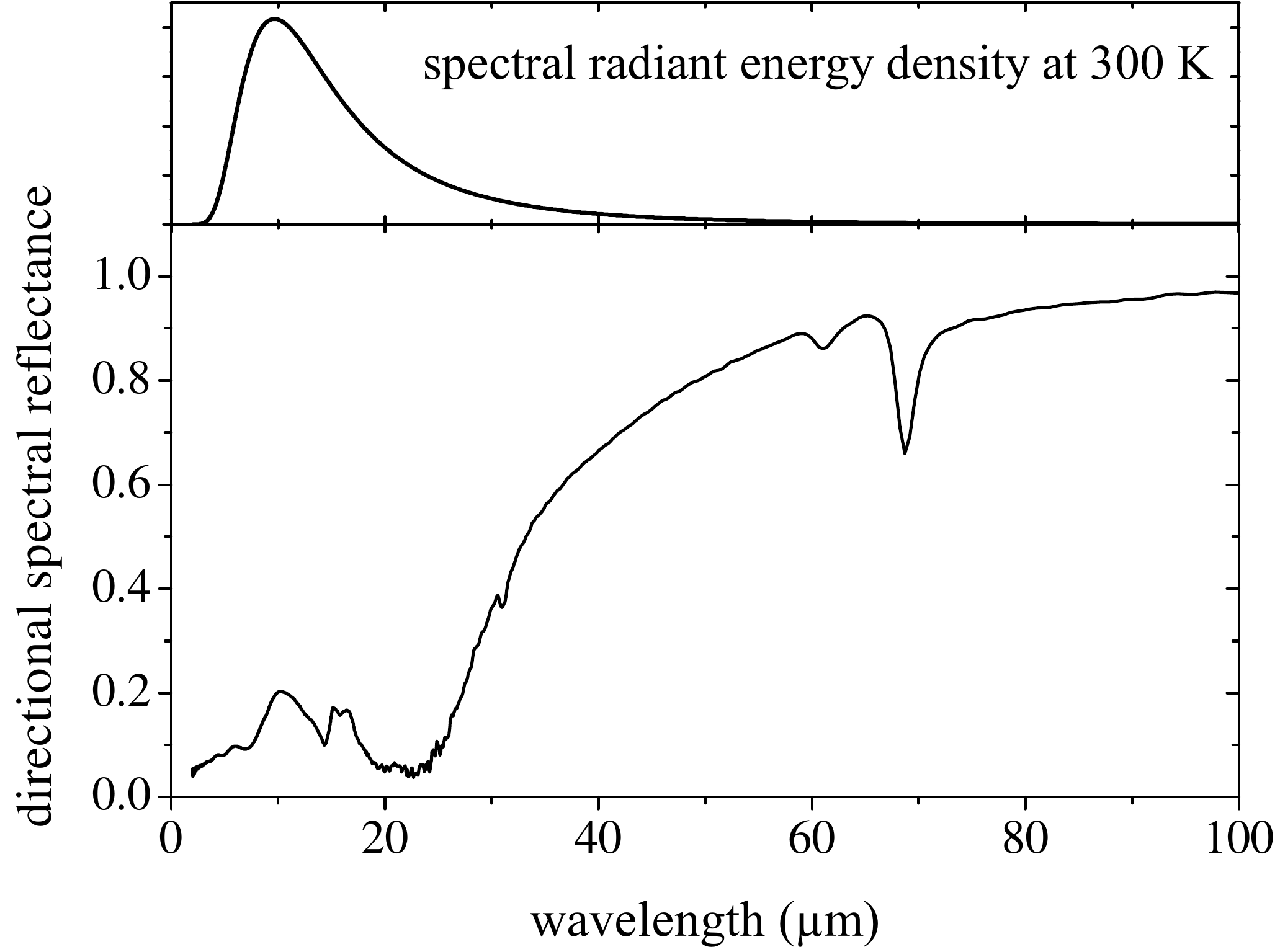}
\caption{Directional spectral reflectance (12$^{\circ}$/12$^{\circ}$) of copper(II) oxide covered copper sample and the distribution of the spectral radiant energy density of a blackbody.}
\label{fig:cuo}
\end{figure}

To minimize and evaluate this disturbance we model the radiation field at the position of the atoms by two contributions.
The first contribution is due to the room temperature BBR irradiating the atoms in direct line 
of sight. The corresponding frequency shift is
\begin{equation}
\Delta\nu^{\rm (dir)}_{\rm BBR}\approx \frac{\Theta}{4\pi} \times 2.354\:{\rm Hz},
\label{dir}
\end{equation}
with the solid angle $\Theta$ of the orifice seen from the position of the atoms.

The second contribution is due to room temperature BBR, which indirectly irradiates the atoms after being reflected from the cavity walls, and BBR emitted from the cold cavity itself.

We estimate the energy density $\rho^{\rm (cav)}$ by looking at the spectral energy flux and setting the total flux into and out of the cavity to be the same.

The relevant fluxes are 
(a) incoming room temperature blackbody radiation $\Phi^{\rm (RT)}(\nu) $, 
(b) blackbody radiation from the inner surface of the cavity $\Phi^{\rm (LN)}(\nu)$, 
(c) absorption at the inner surface of the cavity $\Phi^{\rm (abs)}(\nu)$ and 
(d) leakage out of the orifice $\Phi^{\rm (leak)}(\nu)$.

\begin{figure}[t]
\centering
\includegraphics[width=0.8\columnwidth]{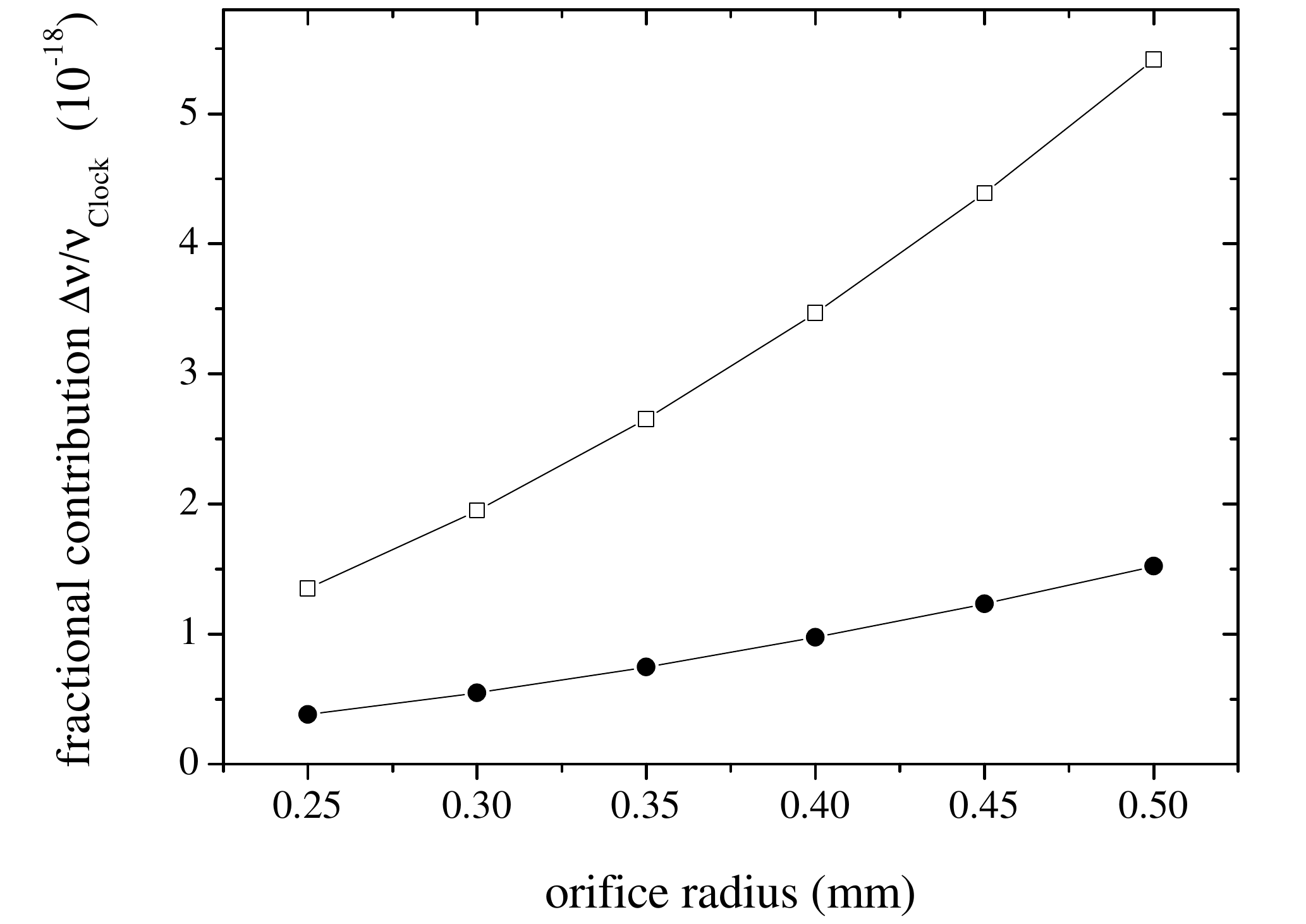}
\caption{Estimated deviation from the BBR shift at $T=$~77~K due to incoming room temperature BBR $T_{\rm RT}=$~300~K as function of the orifice radius. The deviation due to direct exposure of the atoms (full dots) for a distance of 15~mm form the orifice, and the deviation due to indirect radiation (open squares) are shown in parts of the Sr clock frequency. The calculation was performed according to the model presented in the text, for a sphere radius $R=$~10~mm.}
\label{fig:eqenden}
\end{figure}

For simplicity we consider a spherical cavity with a sphere radius $R$
and an orifice radius $r$. 
We assume that the room temperature radiation enters the cavity through the orifice isotropically from a solid angle of $2\pi$, thus neglecting any effects of a finite length orifice. 
Further we assume diffuse reflection on the 
inner walls, having a spectral absorption coefficient $a(\nu)$ and consider
$\rho^{\rm (cav)}$ as being spatially homogeneous and isotropic. With this we obtain 
for the energy rates:
\begin{eqnarray}
\Phi^{\rm (RT)}(\nu) &=& \pi r^2 c\rho(T_{\rm RT},\nu)/4 \nonumber \\
\Phi^{\rm (LN)}(\nu) &=& \pi (4R^2-r^2)\:a(\nu)\:c\rho(T_{\rm LN},\nu) /4 \nonumber \\
\Phi^{\rm (abs)}(\nu) &=& -\pi (4R^2-r^2)\:a(\nu)\:c\rho^{\rm (cav)}(\nu) /4 \nonumber \\
\Phi^{\rm (leak)}(\nu) &=& -\pi r^2 c\rho^{\rm (cav)}(\nu)/4. 
\label{eq:rates}
\end{eqnarray}
With $\rho(T,\nu)$ being the 
spectral radiant energy density for blackbody radiation 
\begin{equation}
\rho (T,\nu)= \frac{8\pi h\nu^3}{c^3}\frac{1}{e^{h\nu/k_B T}-1}
\label{eq:bb}
\end{equation}
at room temperature $T_{\rm RT}$ and liquid nitrogen temperature $T_{\rm LN}$, respectively.

From this we get for the equilibrium spectral energy density inside the cavity 
\begin{equation}
\rho^{\rm (cav)}(\nu) = \frac{(4R^2-r^2)\:a(\nu)\: \rho(T_{\rm LN},\nu) + r^2\rho(T_{\rm RT},\nu)}
{(4R^2-r^2) a(\nu) +  r^2}.
\label{eq:enden}
\end{equation}
For a small orifice the influence of the room temperature is approximately $\Theta/4\pi\cdot1/a(\nu)$. A reflectivity of 50\% would lead to an effect that is twice as big as from the direct contribution (c.f. Eq.~\ref{dir}). In general, small absorption coefficients have to be avoided. To be more specific we have modelled a spherical cavity of $R=$~10~mm coated with a copper(II) oxide (CuO) layer. We have measured the directional spectral reflectance $R(\nu)$ of a CuO layer, obtained by thermal oxidation of a copper sample that resists heating to 550~K and cooling to 77~K. From the measurement (Fig.~\ref{fig:cuo}) at an incident angle of 12$^{\circ}$ (given by the instrument) we deduce the absorption coefficient as $a(\nu)=1-R(\nu)$. As diffuse reflection was ignored, this gives an upper limit for $a(\nu)$.

With this data using Eq.~\ref{eq:enden} the integrated energy density and with Eqs.~\ref{eq:bbrSstat} and \ref{eq:efield} the corresponding BBR shift was calculated and is shown in Fig. \ref{fig:eqenden} in comparison to the shift from the direct contribution.

\begin{figure}[t]
\centering
\includegraphics[width=0.9\columnwidth]{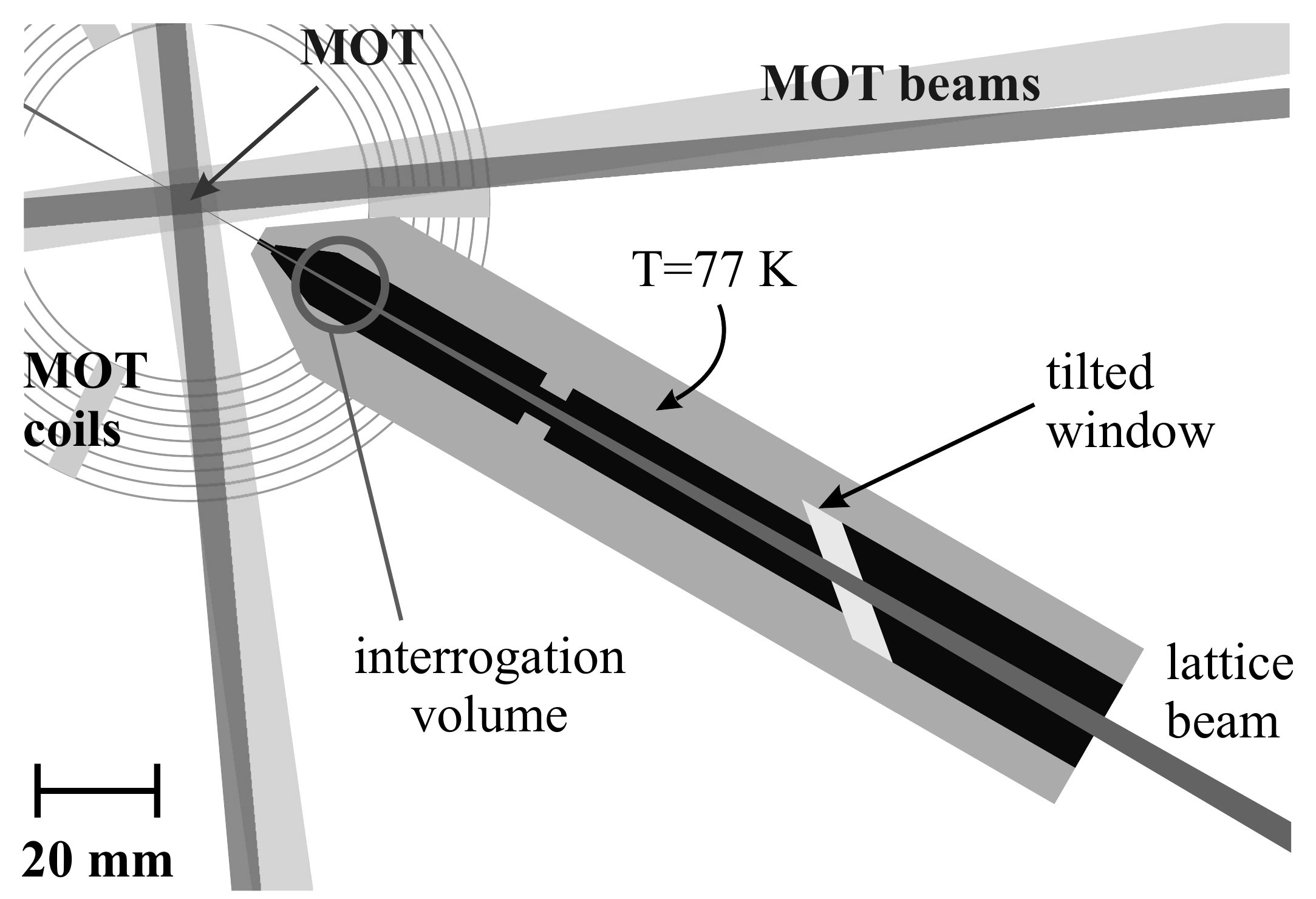}
\caption{Layout of the proposed setup with the cold copper structure, the overlapped blue and red MOT beams and the MOT coils.}
\label{fig:coldfinger}
\end{figure}

Both influences are proportional to the area of the orifice which thus should be as small as possible. 
Due to the spatial restrictions of our setup (Fig.~\ref{fig:coldfinger}) the cold environment cannot be closer to the MOT than about 15~mm. The atoms will be brought into the cold environment by moving the focus of the lattice beam from the MOT region into the cavity. The orifice radius can have its smallest value if it is located at the half transport distance, which is thus at least 30~mm. 
The orifice radius should be larger than twice the $1/e^2$ beam radius of the lattice and the clock laser beam. This leads to a suitable orifice radius of between 0.35 and 0.50~mm. 

Due to the spatial limitations we will use a cylindrical copper structure (Fig.~\ref{fig:coldfinger}) with an inner diameter of about 10~mm and a length of 160~mm. An aperture separates the interrogation volume from the rear part. The inside will be covered by a CuO layer. Compared to the interrogation volume the sphere ($R$~=~10~mm) of the above model has a smaller surface area and thus leads to a conservative estimation of the impact of the diffuse radiation.
To reduce the solid angle of the incoming radiation and to suppress waveguide effects from grazing incidence in the orifice we will use a baffle structure, which we realize by a thread.

The rear end will be closed by a tilted window that is transparent for the lattice laser light but absorbs BBR.
The reflected light is shielded by the aperture, to avoid local hotspots in the interrogation volume. This geometry constitutes a good black body, in which the reflectivity of the window is not important \cite{pro04}.

Nevertheless the lattice laser light will heat the rear volume. Firstly, the heating of the walls by any reflected light can be calculated  with a thermal conductivity of 20~${\rm W/(m\: K)}$ for the CuO layer of 100~${\rm\mu m}$ thickness covering a surface of about  $2\times 10^{-3}\: {\rm m^2}$ and a total power of scattered light of 25~mW  to result in an increase of the average surface temperature of less than 70~${\rm\mu K}$. 
Secondly, absorption of laser light will heat the window. We assume 25~mW of absorbed power, with a thermal conductivity of the sapphire window of 400~${\rm W/(m\: K)}$, it causes a temperature change at the laser spot of less than 20~mK. 
All these effects combined change the effective temperature of of both the rear and the interrogation volume by much less than the required temperature uncertainty of $U(T)=1$~K.

\section{Moving Lattice}
\label{sec:movlat}
As discussed above we need to transport ultracold atoms in an optical lattice from a MOT region to a controlled environment, separated by up to 50~mm. The lattice laser beam is focused to a beam waist radius of $w_0=$~$62\ \mu$m with a Rayleigh range of 15~mm in order to meet the experimental constraints. For short transport distances it is possible to create a moving lattice by detuning the frequency of one of the interfering laser beams with respect to the other beam. But for distances much larger than the Rayleigh range the required higher laser power imposes technical challenges. Generally, a higher laser power also increases the uncertainty of the clock transition due to imperfect knowledge of the used magic wavelength and corresponding hyperpolarizability.

We can circumvent such problems by realizing the transport in straight forward manner: We move all optics for the optical lattice synchronously and thus the interference pattern as it is. In this way, the captured atoms are dragged along and never leave the waist region of the lattice laser beams. In experiments with a test setup we look at possible degradation of the lattice due to, both, direct and parametric heating due to vibrations and loss of overlap.

In the moving lattice setup the optics are mounted on two air bearing translation stages, which will be placed on both sides of the vacuum chamber. On one side the lattice light is emitted from a commercial polarization maintaining large mode area (LMA) fiber, which allows for higher transmitted power than a conventional fiber before Brillouin scattering becomes an issue for light shifts of the clock. The light emitted from the fiber is collimated, passes through polarization optics, and is focused with an achromat of $f=400$~mm to a waist radius of 
$w_0= 62\ \mu$m. On the other side the light is collimated again and retro-reflected by a mirror back into the fiber. For our test experiments we extract the light that comes all the way back through the fiber with a Faraday isolator and record its power with a photo diode.

To maintain the interference pattern during the transport the two waists need to stay overlapped. A radial mismatch $\Delta r$ of the two waist positions, if not too large, leads to an increased anharmonicity and decreased curvature of the lattice potential in the radial direction and also to a reduction of the trap depth. A mismatch of less than half a beam waist reduces the lattice depth at the center to about 88\% and further out saddle points appear between adjacent lattice sites 5\% below the maximum barrier height.

To ensure a good overlap of the two waists a straight motion of the stages with low dynamic and static tilting and a good alignment of the two stages to one another is crucial. Thus we are using air bearing stages, synchronized by a real time control.

The air bearing stages are mounted on adjustment frames. Each frame allows for adjustment of four degrees of freedom: two transverse translations and two angular alignments.
In the present test setup the mutual alignment of motional directions is better than 0.1~mrad. The mutual static tilting is less than $\mathrm{\pm15\: \mu rad}$ over 100~mm and the dynamic tilting of each slide at maximum acceleration is less than $\rm{\pm16\: \mu rad}$ (pitch) and $\mathrm{\pm 5\: \mu rad}$ (yaw) as measured with an autocollimator. Especially the dynamic pitch is roughly proportional to the acceleration. Moving with maximum acceleration of 1~g and maximum speed of 300~mm/s allows for a 100~mm travel in less than 400~ms.

\begin{figure}[t]
\centering
\includegraphics[width=\columnwidth]{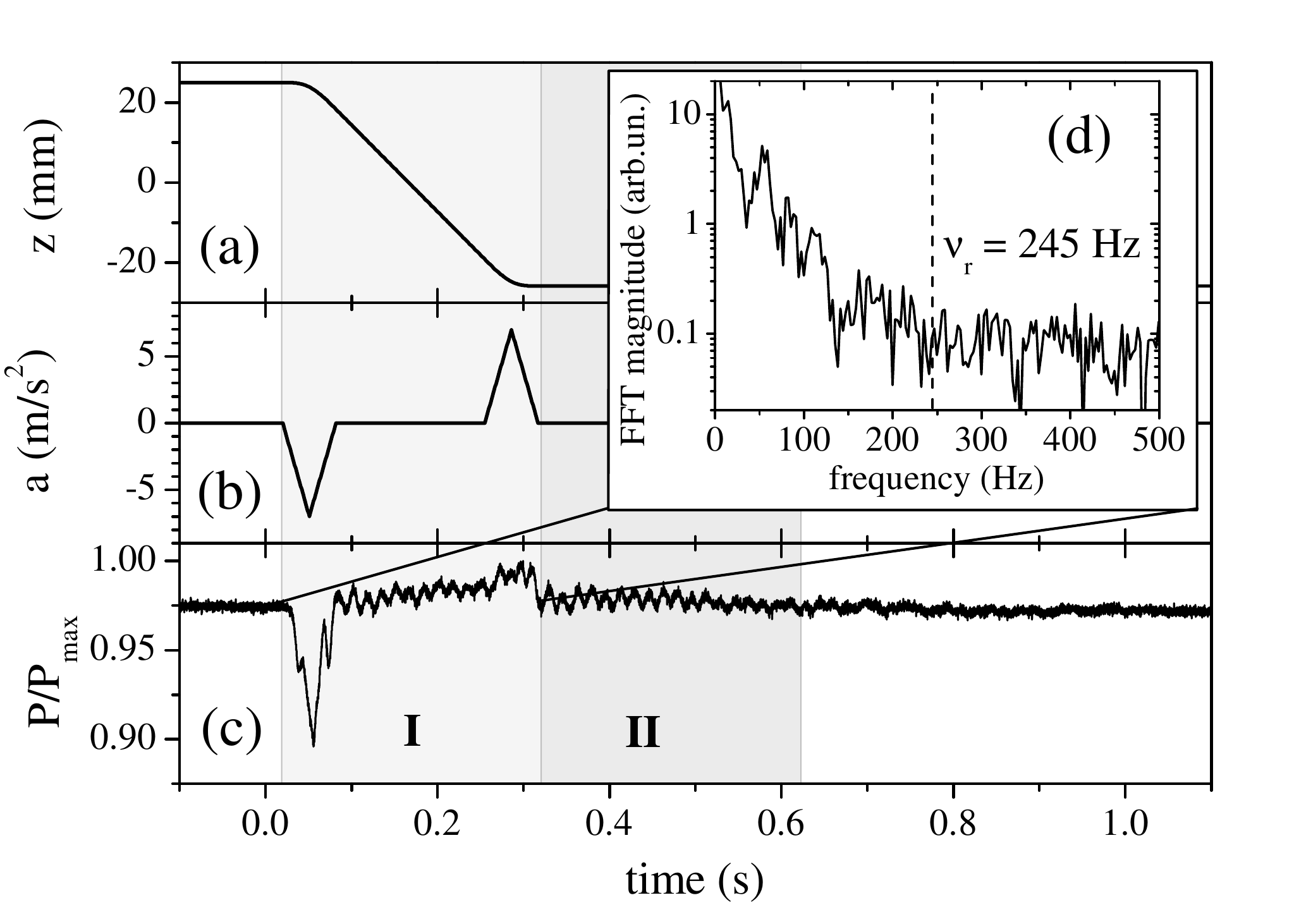}
\caption{Synchronous movement of both air bearing stages over 50~mm within 300~ms (Region I), with max acceleration $a_{max}=7\: {\rm m/s^2}$. The position is shown in (a), the acceleration in (b) and the normalized signal of the retro-reflected light is shown in (c). The inset (d) shows the frequency spectrum of the fluctuations during the movement, in Region I. The radial trap frequency is indicated by the dashed line.}

\label{fig:syn}
\end{figure}

Figure~\ref{fig:syn} shows a translation over 50~mm within 300~ms. The acceleration is linearly ramped up and down, with a maximum value of 7~m/s${\rm^2}$ (b). The power of the retro-reflected light $P$, shown in graph (c), serves as indicator for the overlap of the two waists. The retro-reflected beam is hitting the fiber tip with the same fractional mismatch as the lattice waists.
Since the beam parameters of the retro-reflected beam are matching those of the fiber, the back coupled power in dependency of the mismatch is a convolution of the Gaussian beam profile with itself and as such a Gaussian function of a width $\sqrt{2}\:w_0$:

\begin{equation}
P(\Delta r) = P_{\rm max}\:e^{-\Delta r^2/w_0^2},
\label{eq:mis}
\end{equation}
which allows to calculate the mismatch $\Delta r$. We verified experimentally by tilting the retro reflecting mirror that this equation is applicable, and no correction due to loss in beam quality is necessary.

The initial signal drop in Fig.~\ref{fig:syn}c during the acceleration phase corresponds to a relative radial movement of the two waists by 11~$\rm\mu$m. During the following phase of constant speed the slight position dependent signal change corresponds a movement of 2 to 3~$\rm\mu$m due to static i.e. position dependent tilting. It is about linearly dependent on the position and it becomes more important in the extreme positions of the stages. The change in signal during deceleration is smaller than during the acceleration. Here both waists are shifted in the opposite direction as during acceleration, and thus the initial mismatch is reduced. The signal increase corresponds to a change of the mismatch $\Delta r$ of about 10~$\rm\mu$m. This change of the mismatch is about the same as during acceleration. The difference of the signal during acceleration and deceleration in Fig.~\ref{fig:syn}c is caused by the different slope of the sensitivity function (Eq.~\ref{eq:mis}).

Since the back-coupled signal is correlated to the mismatch of the foci, it can be used to calculate the change in lattice potential depth $U$ as $U/U_0=\sqrt{P/P_{\rm max}}$.
The maximum signal drop in Fig.~\ref{fig:syn}c corresponds to a mismatch of the lattice foci of about 21~$\rm{\mu m}$ or a third of the waist radius respectively, reducing the lattice depth by 5\%. The size of the mismatch is consistent with the dynamic tilting measured and the expected propagation of its impact according to geometrical optics.

During motion beam pointing and lattice depth fluctuations occur that can lead to heating of the trapped atoms. Direct heating due to pointing fluctuations at the trap vibration frequencies and parametric heating by fluctuations of the trap potential curvature at twice the trap vibration frequencies is possible \cite{sav97}.
In our moving lattice setup with 600~mW laser power, which is effectively lowered due to reflections on the windows of the vacuum chamber by factor 0.86, we obtain a lattice potential depth $U\approx 24 \: \mathrm{\mu K} \times k_{B}$ and axial and radial vibration frequencies of $\nu_z=83$~kHz and $\nu_r=245$~Hz respectively.

The mechanical vibrations present during the translation (Fig.~\ref{fig:syn}c, phase I) have significant amplitudes below 150~Hz, as shown in the inset of Fig.~\ref{fig:syn}d. But since the radial trap frequencies are above 200~Hz, these fluctuations are not critical. At the end position (II) the main frequency contributions of the vibrations are at 21~Hz and 60~Hz. 

In the final setup the clock laser beam will be overlapped with the lattice beam through a dichroic mirror that will be used to retro reflect the lattice beam. The clock laser phase will be stabilized relative to this mirror and such the vibrations present in Region II should not perturb the clock interrogation.

\section{Conclusion}

The blackbody radiation emitted by the room temperature environment is causing a significant obstacle for improving optical clocks based on neutral $^{87}$Sr in an optical lattice. In this paper, we have discussed details of two experiments both aiming at a reduction of the BBR influence on the clock accuracy. One experiment is to improve the precision of the difference of the static polarizabilities of the two clock states by a factor of 20. This would improve the BBR correction for room temperature environments such that the imperfect knowledge the static polarizabilities will not limit the accuracy down to a level of $4~\times~10^{-18}$, which corresponds to the uncertainty caused by a temperature uncertainty of 0.05~K.

Our second approach is to interrogate the atoms in a cryogenic environment. We have presented a design of a cryogenic blackbody cavity that will allow us to reduce BBR effects to a level of a few times~$10^{-18}$. The effect of the emissivity of the cold cavity surface and of the hole for atoms to enter and leave this cavity -- in other words the deviations of the cold cavity from a perfect black body -- have been considered. In order to perform the two discussed experiments we need to transport the ultracold ensemble of atoms from the cooling and trapping region into controlled environments. We have presented our setup that will allow to transport such atoms in a mechanically moving optical lattice over 50~mm within 300~ms. 

 We conclude that the possibility of transporting atoms for optical clocks over significant distances may open the way to several new experiments needed for the evaluation of clock uncertainties and for running clocks on ensembles of neutral atoms with unprecedented accuracy -- not only for $^{87}$Sr.


\section*{Acknowledgment}

The authors would like to thank Mikko Merimaa for setting up the stabilization system for our 461~nm cooling laser as well as Christian Monte and Matthias Kehrt for the reflectivity measurements. The support by the Centre of Quantum Engineering and Space-Time Research (QUEST), the European Community's ERA-NET-Plus Programme (Grant No.~217257), and by the ESA and DLR in the project Space Optical Clocks is gratefully acknowledged.



%


\bibliographystyle{unsrt}
\bibliography{O:/4-3/4-3-Alle/Papers/TeXBib/texbi431}

\end{document}